\documentclass[aps,twocolumn,showpacs]{revtex4}
\usepackage{graphics}
\usepackage{epsfig}

\renewcommand{\Re}{\mathrm{Re}}

\begin{document}

\title{Control of unstable steady states by time-delayed feedback methods}

\author{P. H\"ovel} 
\email{phoevel@physik.tu-berlin.de}
\author{E. Sch\"oll}
\affiliation{Institut f\"{u}r Theoretische Physik, TU Berlin, Hardenbergstra{\ss}e 36, D-10623 Berlin,
Germany}

\date{\today}

\begin{abstract} 
We show that time-delayed feedback methods, which  have successfully been used to control unstable periodic ortbits, 
provide a tool to stabilize unstable steady states. We present an analytical investigation of the feedback
scheme using the Lambert function and discuss effects of both a low-pass filter included in the control loop and
non-zero latency times associated with the generation and injection of the feedback signal.
\end{abstract}


\maketitle

\section{Introduction}
Starting with the work of Ott, Grebogi and Yorke \cite{OTT90} a variety of
methods for controlling unstable and chaotic systems
have been developed in the last fifteen years and applied to various real
systems in physics, chemistry, biology, and medicine \cite{SCH99c,BOC00,GAU03}.
Pyragas \cite{PYR92} introduced a time-delayed feedback scheme that
stabilizes unstable periodic orbits (UPO) embedded in a chaotic attractor by
constructing a control force from the difference of the current state to the
state one period in the past. This method is known as \textit{time-delay
autosynchronization} (TDAS) and was improved by Socolar \textit{et al.}
\cite{SOC94} by considering multiple delays in form of an infinite series
(\textit{extended TDAS} or ETDAS) or an average of $N$ past iterates
(\textit{N time delay autosynchronization} or NTDAS)
\cite{SOC98} or coupling matrices (\textit{generalized ETDAS} or GETDAS)
\cite{HAR04}. In parallel
to the control of UPOs, the stabilization of unstable steady states (USS) became
a field of increasing interest.

One of the methods to control an USS introduced by Bielawski \textit{et al.} uses the derivative of the
current state as source of a control force \cite{BIE93}. It can be shown, however, that this \textit{derivative
control} is sensitive to high frequency oscillations \cite{CHA98} and thus not robust in the presence of noise. Another
control scheme was given by calculating the difference of the current state to a low-pass filtered version \cite{PYR04}.

Although the effects of time-delayed feedback schemes on the stability of periodic orbits are understood to large
extend \cite{JUS97,JUS99,PYR02,LOE04,BAL05}, much less is known in the case of a fixed point. There are some results
discussing the application of the ETDAS control method \cite{PYR95a} and numerical simulations of Chua's circuit
\cite{AHL04}, but a detailed theoretical investigation is still missing.

The purpose of this paper is the analytical and numerical study of
the TDAS method, which was originally invented to control
unstable periodic orbits \cite{PYR92}, in application to unstable fixed
points, including latency and filtering effects.
The paper is organized as follows. In
Sec.~\ref{sec:system} we will introduce the system's equations and the control force. In Sec.~\ref{sec:shape} we
will investigate the domain of control in dependence on time delay and feedback gain and present analytical solutions of
the characteristic equation using the Lambert function. Further, we will
consider effects of non-zero latency times
and additional low-passs filtering in Sec.~\ref{sec:latency} and Sec.~\ref{sec:low}, respectively. 

\section{\label{sec:system}Control by time-delayed feedback}
We consider a general dynamic system given by a vector field ${\bf f}$:
\begin{eqnarray}
        \label{eqn:dyn0}
                \dot{\bf x} = {\bf f}({\bf x})
\end{eqnarray}
with an unstable fixed point ${\bf x}^*$ given by ${\bf f}({\bf x^*})=0$.
The stability of this fixed point is obtained by linearizing the vector field
around ${\bf x}^*$. Without loss of generality, let us assume ${\bf x}^*=0$.
In the following we will consider the generic case
of an unstable focus for which the linearized equations in center manifold
coordinates $x, y$ can be written as
\begin{eqnarray}
        \label{eqn:dyn1}
        \dot{x} &=& \lambda \, x + \omega \, y \\
        \dot{y} &=& -\omega \, x + \lambda \, y,  \nonumber
\end{eqnarray}
where $\lambda$ and $\omega$ are positive real numbers. They may be viewed
as parameters governing the distance from the instability threshold,
e.g., a Hopf bifurcation of system (1), and the intrinsic eigenfrequency,
respectively.
For notational convenience, Eq.~(\ref{eqn:dyn1}) can be  rewritten as
\begin{eqnarray}
        \label{eqn:dyn2}
        \dot{\bf x}(t) = {\bf A} \; {\bf x}(t).
\end{eqnarray}
The eigenvalues $\Lambda_0$ of the matrix ${\bf A}$ are given by
$\Lambda_0 = \lambda \pm i \omega$, so that for $\lambda > 0$ and
$\omega \neq 0$ the fixed point is indeed an unstable focus.
A vanishing imaginary part, i.e., $\omega=0$, would correspond
- in the case of a UPO - to an orbit without torsion
for which TDAS fails \cite{JUS97}. We note that the same holds for USSs and
therefore we restrict our investigation to
$\omega \neq 0$.

We shall now apply time-delayed feedback control \cite{PYR92} in order to
stabilize this fixed point:

\begin{eqnarray}
	\label{eqn:uss}
        \dot{x}(t) &=& \lambda \, x(t) + \omega \, y(t) - K [x(t) - x(t-\tau)]\\
        \dot{y}(t) &=& -\omega \, x(t) + \lambda \, y(t) - K [y(t) - y(t-\tau)],
        \nonumber
\end{eqnarray}
where the feedback gain $K$ and the time delay $\tau$ are real numbers.
The goal of the control method is to change the sign of the real part of the
eigenvalue.

Since the control force applied to the i-th component of the system
involves only the same component, this control scheme is called diagonal
coupling \cite{BEC02}, which is suitable for an analytical treatment.
Note that the feedback term vanishes if the USS is stabilized since
$x^*(t-\tau)=x^*(t)$ and $y^*(t-\tau)=y^*(t)$ for all $t$,
indicating the non-invasiveness of the TDAS method.

\begin{figure}[t] 
\epsfxsize= \linewidth
\epsfbox{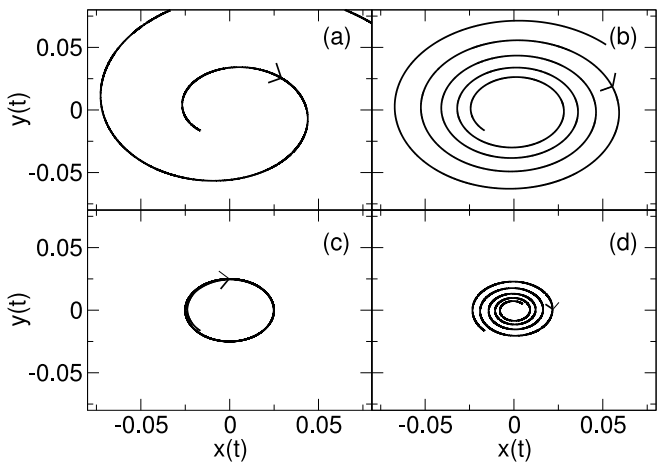}
\caption{\label{fig:figure1}Control of an unstable focus
 with $\lambda = 0.5$ and $\omega = \pi$ in the configuration space for
different values of the feedback gain $K$. Panels (a), (b), (c), and (d) correspond to $K=0, 0.2, 0.25,$ and $0.3$,
respectively. The time delay $\tau$ of the TDAS control scheme is chosen as
$1$, corresponding to $\tau = T_0/2 = \pi/\omega$.
}
\end{figure}
Figure~\ref{fig:figure1} depicts the dynamics of the controlled unstable focus
($\lambda = 0.5$ and $\omega = \pi$) in
the $x$-$y$ plane for different values of the feedback gain $K$.
Panels (a) through (d) correspond to increasing $K$. The time delay
of the TDAS control scheme is chosen as $\tau=1$ in all panels. Panel (a)
displays the case of the absence of control, i.e., $K=0$, and shows that the
system is an unstable focus exhibiting undamped oscillations on a timescale
$T_0\equiv 2\pi/ \omega=2$. It can be seen
from panel (b) that increasing $K$ reduces the instability. The system diverges
more slowly to infinity indicated by the
tighter spiral. Further increase of $K$ stops the unstable behavior
completely and produces periodic motion, i.e., a center [see panel
(c)]. The amplitude of the orbit depends on the initial conditions, which are
chosen as $x=0.01$ and $y=0.01$. For even
larger feedback gains, the trajectory becomes an inward spiral and thus
approaches the fixed point, i.e., the focus.
Hence the TDAS control scheme is successful.

An exponential ansatz for $x(t)$ and $y(t)$ in Eq.~(\ref{eqn:uss}), i.e.,
$x(t) \sim \exp(\Lambda t), \; y(t) \sim
\exp(\Lambda t)$, reveals how the control force modifies the eigenvalues of
the system. The characteristic equation becomes
\begin{eqnarray}
        [\Lambda + K \left( 1 - e^{- \Lambda \tau}\right)-\lambda]^2+\omega^2=0.
\end{eqnarray}
so that the complex eigenvalues $\Lambda$ are given in the presence of a
control force by the implicit equation
\begin{eqnarray}
	\label{eqn:characteristic} 
	\lambda \pm i \omega &=& \Lambda + K \left( 1 - e^{- \Lambda \tau}\right).
\end{eqnarray}
Using the Lambert function $W$, which is defined as the inverse function of
$g(z) = z e^z$ for complex $z$
\cite{WRI49,WRI55,BEL63,HAL71,ASL03}, Eq.~(\ref{eqn:characteristic}) can be solved analytically
\begin{eqnarray}\label{eqn:lambert} 
	\Lambda \tau &=& W\left(K \tau e^{- (\lambda \pm i \omega) \tau + K \tau}\right) + (\lambda \pm i \omega) \tau -
K \tau.
\end{eqnarray}

\begin{figure}[t] 
\epsfxsize=\linewidth
\epsfbox{figure2a.eps}\\
\epsfxsize=\linewidth
\epsfbox{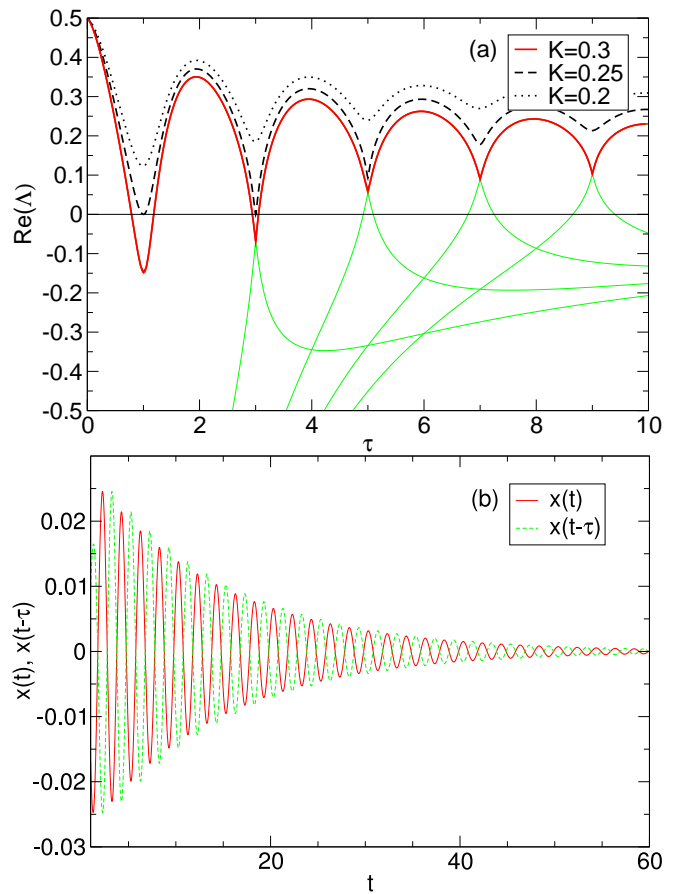}
\caption{\label{fig:figure2}(Color online)(a) Largest real part of the complex eigenvalues
$\Lambda$ vs. $\tau$ for $\lambda = 0.5$ and $\omega = \pi$ for different $K$.
Some lower eigenvalues are also displayed for $K=0.3$ (green online).
(b) Time series of the $x$-component of the unstable focus: The solid line
(red online) corresponds to $x(t)$, the dashed line (green online)
to the delayed $x$ component $x(t-\tau)$ with $\tau =1$. The parameters of
the unstable focus and the
control scheme are as in panel (d) of Fig.~\ref{fig:figure1}.}
\end{figure} 
Panel (a) of Fig.~\ref{fig:figure2} shows the dependence of the largest real part of the complex eigenvalues $\Lambda$
upon
the time delay $\tau$ according to Eqs.~(\ref{eqn:characteristic}),
(\ref{eqn:lambert}) for $\lambda = 0.5$ and $\omega =
\pi$. The solid curve corresponds to a feedback gain of $K=0.3$, the dashed curve to $K=0.25$, and the dotted curve to
$K=0.2$. All curves start at $\Re(\Lambda)=\lambda$ for $\tau=0$, i.e., when
no control is applied to system. For increasing time delay,
the real part $\Re(\Lambda)$ decreases. It can be seen in the case of $K=0.3$
that there exist values of the time delay for which $\Re(\Lambda)$ becomes
negative, and thus the control is successful. The curve for $K=0.25$ shows
the threshold
case where $\Re(\Lambda)$ becomes zero for $\tau=1$, but does not change sign. The TDAS control scheme generates
an infinite number of additional eigenmodes. The corresponding eigenvalues are the solutions of the transcendental
Eq.~(\ref{eqn:characteristic}). The real parts of the eigenvalues all originate
from $-\infty$ for $\tau=0$. Some of these lower
eigenvalues are displayed for $K=0.3$. The different branches of the eigenvalue
spectrum originate from the multiple-leaf structure of the complex Lambert
function. The real part of each eigenvalue branch exhibits a typical
nonmonotonic dependence upon $\tau$ which leads to crossover of different
branches resulting in an oscillatory modulation of the largest real part
as a function of $\tau$.
Such behavior of the eigenvalue spectrum appears to be quite general, and
has been found for various delayed feedback coupling schemes, including
the Floquet spectrum of UPOs \cite{BEC02,JUS03} and applications to
noise-induced motion where the fixed point is stable \cite{JAN03}.

The notch at $\tau=1$ corresponds to Fig.~\ref{fig:figure1}, so that at this value of $\tau$ the solid, dashed, and
dotted curves correspond to panels (d), (c), and (b) of Fig.~\ref{fig:figure1}, respectively. The notches at larger
$\tau$ become less pronounced leading to less effective realization of the TDAS control scheme, i.e., a smaller or no
$\tau$-interval with negative $\Re(\Lambda)$.

In the case of an UPO the optimal time delay is equal to the period of the
orbit to be stabilized. Note that in the case of an USS, however, the time delay is not so obviously related to a
parameter
of the system. We will see in Sec.~\ref{sec:shape} which combinations of the feedback gain
$K$ and the time delay $\tau$ lead to successful control.

Panel (b) of Fig.~\ref{fig:figure2} displays the time evolution of $x(t)$ and its time delayed
counterpart $x(t-\tau)$ in the case of a combination of $K=0.3$ and $\tau=1$ that leads to successul control as in panel
(d) of Fig.~\ref{fig:figure1}. The $x$ component of the control force can be calculated from the difference of the two
curves and subsequent multiplication by $K$. Since $x(t)$ tends to zero in the limit of large $t$ (the system  reaches the focus
located at the origin), the control force vanishes if the system is
stabilized. Thus the control scheme is non-invasive.
Note that the current signal (red online) and its delayed counterpart (green
online) are in anti-phase. This observation will become important in
Sec.~\ref{sec:latency}.

In the following discussion, it is helpful to consider the real and
imaginary part of Eq.~(\ref{eqn:characteristic}) separately in order to gain some analytic information about the domain
of control:
\begin{eqnarray}
	\label{eqn:characteristic_sep} 
	p + K \left[1-e^{-p \tau} \cos(q \tau)\right] &=& \lambda \\
        q +  K e^{-p \tau} \sin(q \tau) &=& \omega \nonumber
\label{eqn:characteristic_Re_Im}
\end{eqnarray}
with $\Lambda = p + i q$.

\section{\label{sec:shape}Shape of the domain of control}
This section is focused on the construction of the domain of control in the $K$-$\tau$~plane. The calculation can
be done analytically for special points by using, for instance, that
$p = 0$ at the threshold of control.
Furthermore, we will present an expansion around the minimal value of $K$
that reveals further details of the shape of the domain of control.

At the threshold of control the sign of the real part $p$ of the exponent
$\Lambda$ changes. Therefore setting $p$ to zero in the real and imaginary parts, respectively, of
Eq.~(\ref{eqn:characteristic_sep}) yields
\begin{eqnarray}
	\label{eqn:K_min1}
	\lambda &=& K \left[1 - \cos(q \tau) \right]
\end{eqnarray}
and \begin{eqnarray}
	\label{eqn:imaginary}
        \omega = q+K \sin(q \tau).
\end{eqnarray} 

Since the cosine is bounded between $-1$ and $1$, the following inequality
follows from Eq.~(\ref{eqn:K_min1})
\begin{eqnarray}
	\label{eqn:K_min2}
	\frac{\lambda}{2}  &\leq& K.
\end{eqnarray}
Thus a minimal value of $K$, $K_{min} = \lambda/2$, for which the control
starts, can be inferred \cite{note1}. It corresponds to
$q \tau = (2n+1) \pi$ for $n = 0, 1, 2, \dots$.

In order to express the values of the time delay $\tau$ that correspond to
the minimal $K$ in terms of the parameters of the uncontrolled system,
it is useful to consider
even and odd  multiples of $\pi$ for $q \tau$, i.e., $q
\tau = 2 n \pi $ and $q \tau = (2 n +1) \pi $ for $n = 0, 1, 2, \dots$.
In both cases, the imaginary part of Eq.~(\ref{eqn:characteristic})
leads to  $q = \omega$. Hence, in the latter case, the time delay $\tau$ for
$K_{min} = \lambda/2$ becomes
\begin{eqnarray}
	\tau &=& \frac{\pi}{\omega} (2 n + 1).
\end{eqnarray} 
The last expression can be rewritten using the uncontrolled eigenperiod
$T_0$ 
\begin{eqnarray}
	\tau &=& T_0 \frac{2 n + 1}{2},
\end{eqnarray} 
where $T_0$ is defined by
\begin{eqnarray}
	T_0 = \frac{2\pi}{\omega}.
\end{eqnarray}
This discussion has shown that $K = \lambda/2$ and $\tau = T_0 (2n+1)/2$
with $n = 0, 1, 2, \dots$
correspond to points of successful control in the $K$-$\tau$~ plane 
with minimal feedback gain. 

For even multiples, i.e., $q \tau = 2 n \pi$ for $n = 0, 1, 2,
\dots$, no control is possible for finite
values of $K$, since
\begin{eqnarray}
	\frac{K-\lambda}{K} &=& \left. \cos(q \tau)\right|_{q \tau = 2 n \pi} \\
	\Leftrightarrow  1-\frac{\lambda}{K} &=& 1,
\end{eqnarray} 
which cannot be satisfied for $\lambda \neq 0$ and finite $K$. Furthermore, Eq.~(\ref{eqn:imaginary}) yields that for
time delays, which are integer multiples of the eigenperiod, i.e., $\tau = T_0 n = 2 \pi n /\omega$ with $n = 0, 1, 2,
\dots$, the control scheme fails for any feedback gain.
Note that this failure appears to be related to the case of torsion free
UPOs, where it has been shown that $\omega \neq 0$ is a necessary condition for control \cite{JUS97}.

Another result that can be derived from Eq.~(\ref{eqn:characteristic}) is a shift of $q$
for increasing $K$. For this, taking the square of the real and imaginary part of Eq.~(\ref{eqn:characteristic}) and
using trigonometrical identities yields
\begin{eqnarray}
	\label{eqn:ImLambda}
	q = \omega \mp \sqrt{(2K-\lambda)\lambda}.
\end{eqnarray} 
Inserting Eq.~(\ref{eqn:ImLambda}) into the real part of
Eq.~(\ref{eqn:characteristic}) leads to an explicit expression for the
dependence of time delay $\tau$ on the feedback gain $K$ at the threshold of
stability, i.e., the boundary of the control domain $p = 0$,
\begin{eqnarray}
	\label{eqn:realchar}
	\frac{K-\lambda}{K}  &=& \cos(q \tau)\\
	\label{eqn:tauK}
	\Leftrightarrow \hspace{0.5cm} \tau(K) &=& \frac{\arccos \left(\frac{K-\lambda}{K}\right)}{\omega \mp
\sqrt{(2K-\lambda)\lambda}}.
\end{eqnarray}
In order to visualize the shape of the domain of control we will investigate how small deviations
$\epsilon >0$ from $K_{min}$, i.e, $K=\lambda/2 + \epsilon$, influence the corresponding values of the time delay
$\tau$. For
this, let $\eta > 0$ be small and $\tau = \frac{\pi}{\omega} (2n +1) \pm \eta$ a small deviation from $\tau$
at $K_{min}$. Inserting the expression for $K$ and $\tau$ into
Eq.~(\ref{eqn:realchar}) yields after some Taylor's expansions
\begin{eqnarray}
	-1 + \frac{4}{\lambda} \epsilon &=& -1 + \frac{1}{2} \left[\omega \eta \mp \frac{\pi}{\omega} (2n +1)
\sqrt{2\lambda} \sqrt{\epsilon}\right]^2 \\
	\Leftrightarrow \hspace{0.5cm}\eta &=& \left[\pm \frac{2\sqrt{2}}{\omega \sqrt \lambda} + \frac{\sqrt{2}
\pi}{\omega^2} (2n +1) \sqrt{\lambda} \right] \sqrt{\epsilon}.
\end{eqnarray} 
This equation describes the shape of the domain of control at the threshold of
stabilization, i.e., $p=0$, near
the minimum $K$ value at $\tau = T_0 (2n+1)/2$ in the $K$-$\tau$ control plane. Small deviations from $\tau$ at
$K_{min}$ are influenced by the square root of small deviations from the
minimum feedback gain.

\begin{figure}[t] 
\epsfxsize=\linewidth
\epsfbox{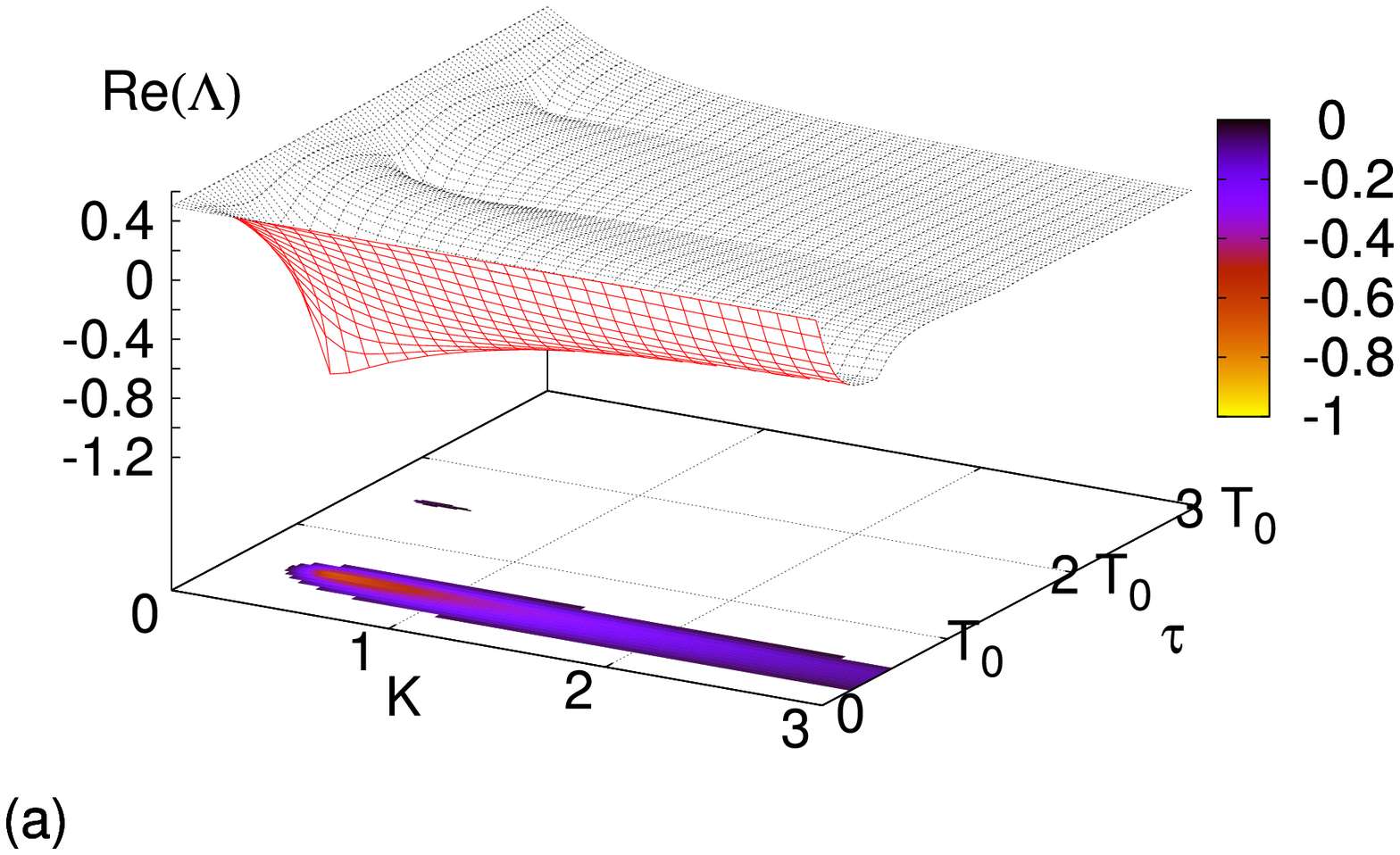}\\
\epsfxsize=\linewidth
\epsfbox{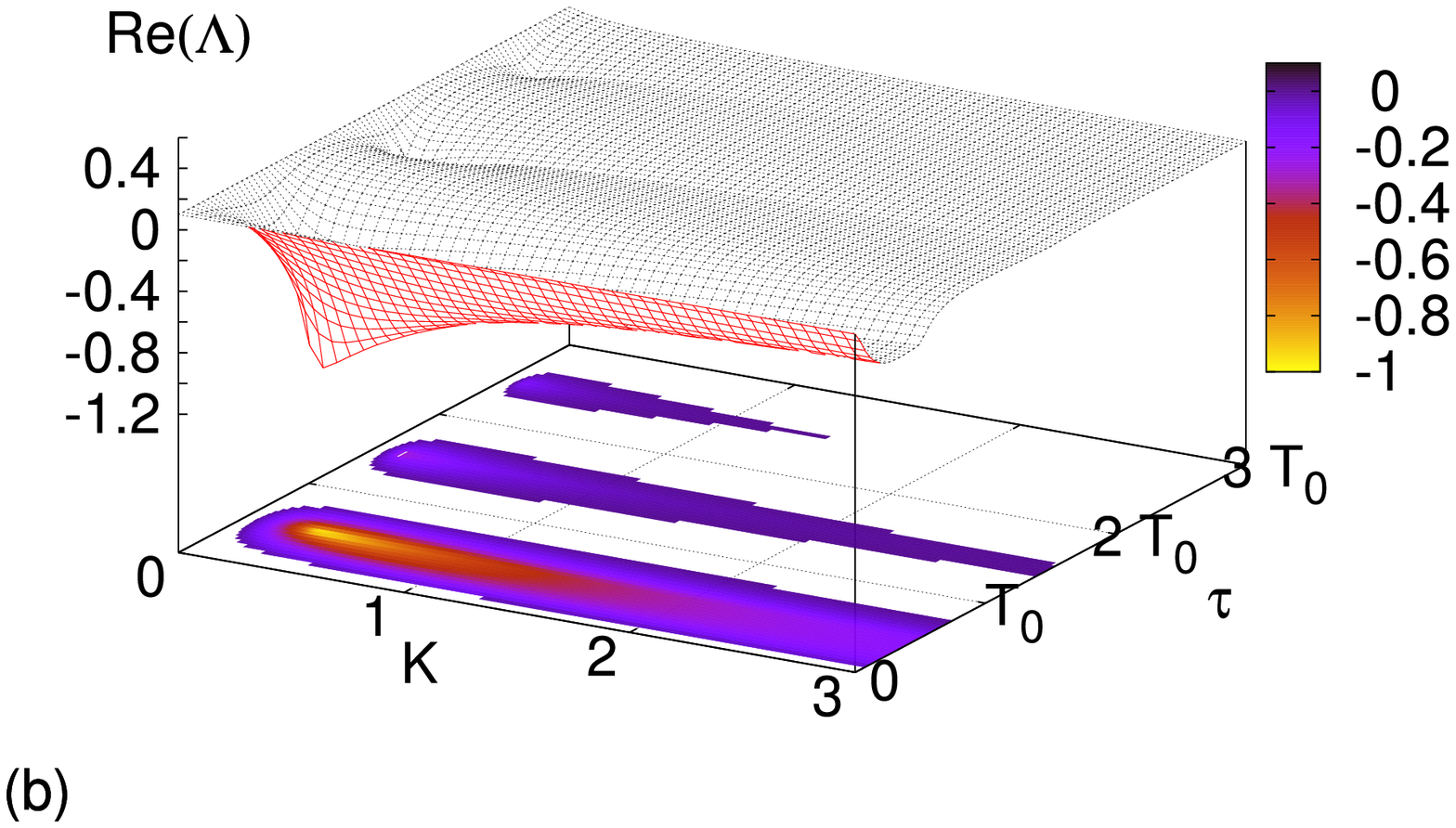}
\caption{\label{fig:figure3} (Color online)
Domain of control in the
$K$-$\tau$~plane and largest real part of the complex eigenvalues
$\Lambda$ as a function of $K$ and $\tau$ according to Eq.~(\ref{eqn:lambert}).
The two-dimensional projection at the bottom shows combinations of $\tau$
and $K$, for which
$\Re(\Lambda)$ is negative and thus the control successful
[Panel (a): $\lambda=0.5$ and $\omega=\pi$, panel (b): $\lambda=0.1$ and $\omega=\pi$].}
\end{figure}
Figure~\ref{fig:figure3} displays the largest real part of the eigenvalues $\Lambda$ in dependence on both the
feedback
gain $K$ and the time delay $\tau$ for $\omega = \pi$ and two different
values of $\lambda$, and summarizes the results of this section.
The values of $\Lambda$ are calculated using the analytic solution (\ref{eqn:lambert}) of
Eq.(\ref{eqn:characteristic}). The two-dimensional projections at the bottom of each plot extract combinations of $K$
and $\tau$ with negative $p$, i.e., successful control of the system. In the absence of a control force, i.e., $K=0$,
the real part of $\Lambda$ starts at $\lambda$. Increasing the feedback gain decreases
$\Re(\Lambda)$. For $K=K_{min}=\lambda/2$, the real part of the eigenvalue reaches $0$ for certain time delays, i.e.,
$\tau = T_0 (2n+1)/2$ with $n = 0, 1, 2, \dots$, and then changes sign. Thus, the system is stabilized. For values of
the feedback gain slighty above the minimum value $K_{min}$, the domain of control shows a square root shape. It can be
seen that for time delays of $\tau = T_0 n$ the largest real part of the eigenvalues remains positive for any feedback
gain. For a smaller value of $\lambda$ (Fig.~\ref{fig:figure3}b), i.e., closer to the instability threshold of the fixed
point, the domains of control become larger.

An example of the combination of minimal feedback gain $K_{min}=\lambda/2$ and corresponding time delay $\tau = T_0 (2 n
+1)/2$, $n = 0, 1, 2, \dots$ is shown in panel (c) of Fig.~\ref{fig:figure1}, where $K=\lambda/2 = 0.25$ and $\tau = T_0
/2 = \pi/\omega = 1$. It describes the control threshold case between stable and unstable fixed point.  

\section{\label{sec:latency}Latency time effects}
In this section we will consider non-zero latency times, which can be
associated with the generation and injection of
the feedback signal \cite{BLA04a}. It has been shown experimentally \cite{SUK97} in the case of an UPO that latency can
have important
effects on the controllability of the system and might limit the success of the time-delayed feedback method. A
theoretical explanation can be found in \cite{JUS99b,HOE03}. Here we will discuss how latency times change the domain
of control in the case of an USS.

The latency time $\delta$ can be included as an additional time delay in the control force of Eq.~(\ref{eqn:uss}),which
then becomes
\begin{eqnarray}
        {\bf F}(t-\delta) = - K \left(
	\begin{array}{c}
                x(t-\delta) - x(t-\tau-\delta)\\
                y(t-\delta) - y(t-\tau-\delta)
	\end{array} 
	\right),
\end{eqnarray}
leading to a characteristic equation similar to Eq.~(\ref{eqn:characteristic})
but with an additional exponential factor
\begin{eqnarray}
	\label{eqn:characteristic_latency} 
	\lambda \pm i \omega &=& \Lambda + K e^{- \Lambda \delta} \left( 1 - e^{- \Lambda \tau}\right)
\end{eqnarray} 
or, separating into real and imaginary parts,
\begin{eqnarray}
	 p + K \left[e^{-p \delta} \cos(q \delta)-e^{-p (\tau+\delta)} \cos{\textbf (}q(\tau+\delta){\textbf )}\right] &=& \lambda \nonumber\\
        q - K \left[e^{-p \delta} \sin(q \delta)-e^{-p (\tau+\delta)} \sin{\textbf (}q (\tau+\delta){\textbf )}\right] &=& \omega \nonumber,
\end{eqnarray}
where $p$ and $q$ denote the real and imaginary part of $\Lambda$, respectively.
\begin{figure}[t] 
\epsfxsize=\linewidth
\epsfbox{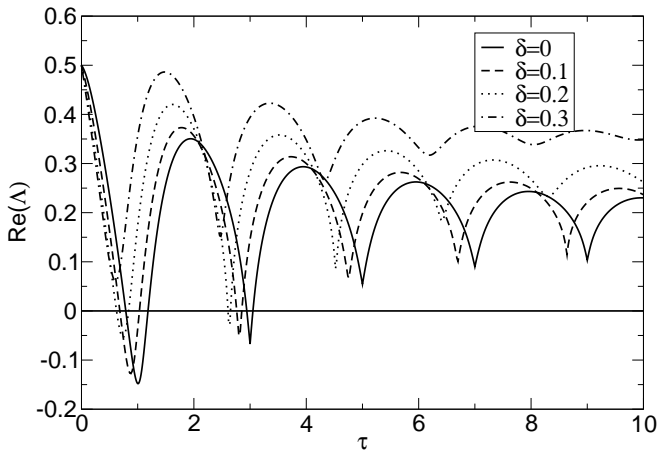}
\caption{\label{fig:figure4}Largest real part of the eigenvalues $\Lambda$ vs. $\tau$ for $\lambda = 0.5$, $\omega =
\pi$, and $K=0.3$ as given by Eq.~(\ref{eqn:characteristic_latency}). The solid, dashed, dotted, and dash-dotted
curves correspond to a latency time of $\delta=0$, $0.1$, $0.2$, and $0.3$, respectively.}
\end{figure}

Figure~\ref{fig:figure4} displays the dependence of the largest real part of the complex eigenvalues $\Lambda$ on the
time delay $\tau$ according to Eq.~(\ref{eqn:characteristic_latency}) for $\lambda = 0.5$ and $\omega = \pi$, and
different values of the latency time $\delta$. The values of the eigenvalues are
calculated by solving Eq.~(\ref{eqn:characteristic_latency}) numerically. The solid, dashed, dotted, and dash-dotted
curves correspond to $\delta=0$, $0.1$, $0.2$, and $0.3$, respectively. The case with zero latency time is also
displayed; it corresponds to the solid curve in Fig.~\ref{fig:figure2}(a).

It can be seen that increasing latency time shifts the minimum of $p$ to smaller values of $\tau$ and reduces the
intervals of $\tau$, for which $p$ is negative, i.e., for which the control is successful. The dash-dotted
curve ($\delta = 0.3$) shows a case where no control is possible since the largest real part of the complex eigenvalues
remains positive for all time delays $\tau$.

\begin{figure}[t] 
\epsfxsize=\linewidth
\epsfbox{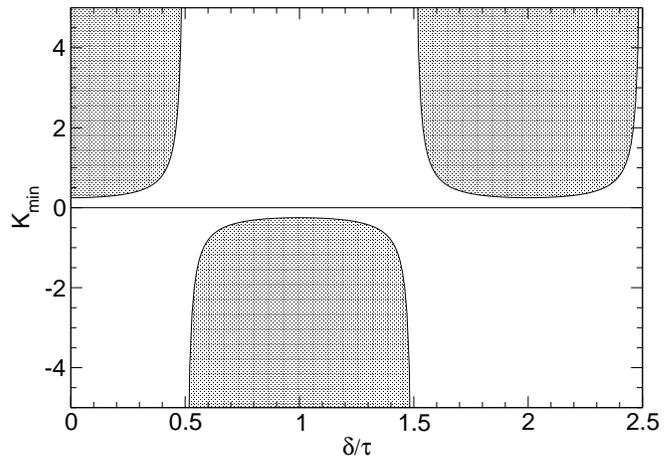}
\caption{\label{fig:figure5}
Minimal feedback gain $K$ vs. relative latency $\delta/\tau$ for $\lambda=0.5$ and $\omega = \pi$ according to
Eq.~(\ref{eqn:K_min_latency}). The shaded areas show the domain of control for suitably chosen $\tau$.}
\end{figure}

Solving Eq.~(\ref{eqn:characteristic_latency}) analytically by using the Lambert function, as for
Eq.~(\ref{eqn:characteristic}), is not possible in the case of non-zero latency times due to the additional exponential
term $\exp(-\Lambda \delta)$. In order to understand the effects of non-zero latency times on the value of the minimal
feedback gain, we evaluate the real part of Eq.~(\ref{eqn:characteristic_latency}) at the threshold of control, i.e.,
$p=0$. It was shown in Sec.~\ref{sec:shape} that $K$ becomes minimal if $q \tau = \pi (2n +1)$
for $n=0,1,2,\dots$ [see Eqs.~(\ref{eqn:K_min1}),(\ref{eqn:K_min2})]. This value of $q\tau$ yields for
non-zero latency times
\begin{eqnarray}
	\label{eqn:K_min_latency}
	K_{min}(\delta) &=& \frac{\lambda}{2 \cos[\pi (2n+1) \frac{\delta}{\tau}]} \geq \frac{\lambda}{2}.
\end{eqnarray}
This shows that non-zero latency times shift the minimal feedback gain $K_{min}$, for which the control scheme is
successful, to larger values. 

The effects of the cosine function in Eq.~(\ref{eqn:K_min_latency}) can be understood by considering
Fig.~\ref{fig:figure2}(b) and Fig.~\ref{fig:figure5}, which shows the dependence of the minimal feedback gain
$K_{min}$ on the latency time for $\lambda=0.5$. Increasing latency time increases the value of $K_{min}$. If $\delta$
becomes larger than half the time delay $\tau$, $K_{min}$ changes sign and control is possible
only for negative $K$ with $K<K_{min}$ and suitably chosen $\tau$. Note that in Fig.~\ref{fig:figure2} the difference
$x(t)-x(t-\tau)$ has to be positive for successful control. For $0.5 \tau < \delta < 1.5 \tau$ both $x(t-\delta)$ and
$x(t-\delta-\tau)$ becomes closer to zero. Therefore in order to achieve control, the feedback gain becomes larger. In
the limit $\delta / \tau \rightarrow 1/2$ the difference $x(t-\delta)-x(t-\tau-\delta)$ vanishes and thus the minimal
feedback gain diverges. For even larger values of $\delta$ the above-mentioned difference changes its sign forcing
$K_{min}$ to do the same. Otherwise the control scheme would generate a force that pulls the system away from the target
fixed point.

\begin{figure}[t] 
\epsfxsize=1.0\linewidth
\epsfbox{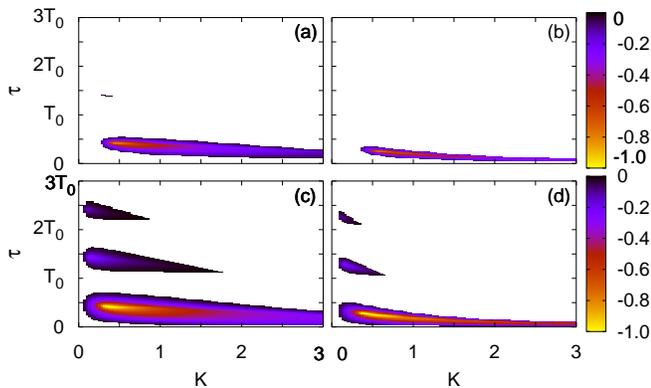}
\caption{\label{fig:figure6} (Color online)
Domain of control in the $K$-$\tau$~plane for different latency times [panels (a) and (c): $\delta=0.1$, panels
(b) and (d): $\delta=0.3$]: The shaded areas indicate combinations of $\tau$ and $K$, for which the largest real part
of the complex eigenvalues $\Lambda$ is negative and thus control is successful. The value of $\Re(\Lambda)$ is
indicated by the greyscale (color online). The parameters of the unstable focus are chosen as $\omega=\pi$ in all
panels and $\lambda=0.5$ in (a), (b) and $\lambda=0.1$ in (c), (d).}
\end{figure}
Figure~\ref{fig:figure6} depicts the domain of control for latency times of $\delta=0.1$ [panels (a) and (c)]
and $\delta=0.3$ [panels (b) and (d)]. The largest real part of the complex eigenvalues $\Lambda$ is shown by greyscale
in the domain. It can be seen that increasing latency times reduce the domain of control. For instance, the small range
at a time delay of $\tau=1.5 T_0$ in (a), where control is possible for $\delta=0.1$, vanishes for $\delta=0.3$ in (b).
Note that non-zero latency times lead to a loss of the symmetry [around $\tau=(2n+1)/2 T_0$ for $n=1,2,\dots$] of the
domain of control (see also the case of zero latency as displayed in Fig.~\ref{fig:figure3}).

\section{\label{sec:low}Low-pass filtering}
It has been found that high frequency modulations of the control signal, due to additional high frequency components in
the signal besides the main frequency, can render the TDAS control method unstable \cite{SCH03a}. As shown in that work,
an additional low-pass filter included in the control loop can overcome this limitation, and UPOs can be stabilized
\cite{SCH03a,SCH04a}. On the other hand, in electronic signal processing the finite response time of the circuit often
imposes unavoidable low-pass filtering, and its effect upon feedback control is not clear. In this section we will show
that a low-pass filter changes the characteristic equation [see Eq.~(\ref{eqn:characteristic})] of the fixed point, and
shifts the minimal feedback gain to larger values. Note that low-pass filtering has been successfully used to stabilize
USSs by generating a control force from the difference of the current state to its filtered counterpart \cite{PYR04}.

The TDAS control force with an additional low-pass filter can be written as
\begin{eqnarray}
	{\bf F}(t) = - K \left(
	\begin{array}{c}
		\overline{x}(t) - \overline{x}(t-\tau)\\
		\overline{y}(t) - \overline{y}(t-\tau)
	\end{array} 
	\right),
\end{eqnarray}
where $\overline{x}$ and $\overline{y}$ denote the filtered versions of $x$ and $y$ defined by
\begin{eqnarray}
	\overline{x}(t) = \alpha \int^t_{- \infty}x(t') e^{- \alpha (t - t')} dt'
\end{eqnarray}
with the cut-off frequency $\alpha$, and analogously for $\overline{y}(t)$.
Equivalently, the convolution integrals can be replaced by 
two additional differential equations such that the
original two-dimensional system becomes four-dimensional
\begin{eqnarray}
        \dot{x}(t) &=& \lambda \, x(t) + \omega \, y(t) - K [\overline{x}(t) - \overline{x}(t-\tau)]\\
        \dot{y}(t) &=& -\omega \, x(t) + \lambda \, y(t) - K [\overline{y}(t) - \overline{y}(t-\tau)] \nonumber\\
        \dot{\overline{x}}(t) &=& - \alpha \overline{x}(t) + \alpha x(t) \nonumber\\
        \dot{\overline{y}}(t) &=& - \alpha \overline{y}(t) + \alpha y(t). \nonumber
\end{eqnarray}
This system of differential equations yields a characteristic equation of the form
\begin{equation}
	\label{eqn:characteristic_low}
        \pm i(\alpha + \Lambda) \omega = \alpha K \left(1 - e^{-\Lambda \tau}\right) - (\alpha + \Lambda) (\lambda
-\Lambda)
\end{equation}
or, equivalently, using $\Lambda = p+i q$
\begin{eqnarray}
	\label{eqn:characteristic_low_sep}
	\alpha (\lambda -p) &=&\alpha K \left[1-e^{-p\tau}\cos(q\tau)\right] - q(q- \omega)\\
	\alpha (\omega - q) &=& \alpha K e^{-p\tau} \sin(q \tau) -\lambda q -p\omega.\nonumber
\end{eqnarray} 
Note that in the limit of large cut-off frequencies, i.e., $\alpha \rightarrow \infty$,
Eqs.~(\ref{eqn:characteristic_low}) and~(\ref{eqn:characteristic_low_sep})  reduce to the characteristic
equations (\ref{eqn:characteristic}) and (\ref{eqn:characteristic_sep}) of Sec.~\ref{sec:system}, respectively.

\begin{figure}[t] 
\epsfxsize=\linewidth
\epsfbox{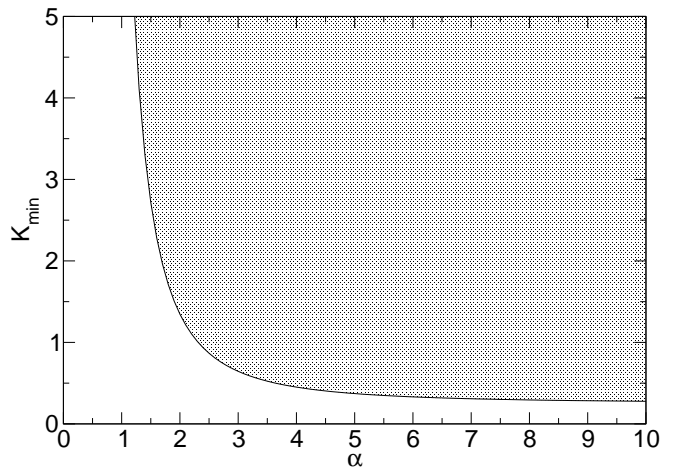}
\caption{\label{fig:figure7}
Minimal feedback gain $K$ vs. cut-off frequency $\alpha$ for $\lambda=0.5$ and $\omega=\pi$
according to Eq.~(\ref{eqn:K_min_lowpass}). The shaded area shows the domain of control.}
\end{figure}

For further investigation of Eq.~(\ref{eqn:characteristic_low}) we shall use
the separation into real and imaginary parts
(\ref{eqn:characteristic_low_sep}). Following the discussion
of Sec.~\ref{sec:shape} by considering even and odd
multiples of $\pi$ as special values for $q\tau$, we obtain
from the imaginary part of
Eq.~(\ref{eqn:characteristic_low}) in the case $q\tau=(2n+1)\pi$
\begin{eqnarray}\label{eqn:tau_alpha}
	\tau = \frac{2n+1}{\omega} \;\pi\; \left(1-\frac{\lambda}{\alpha}\right)
\end{eqnarray} 
with $n=0, 1,2,\dots$. Inserting this into the real part of
Eq.~(\ref{eqn:characteristic_low}) gives an expression for
the minimum value of the feedback gain, for which the control method becomes
successful for appropriately chosen
$\tau$, 
\begin{eqnarray}
	\label{eqn:K_min_lowpass} 
	K_{min}(\alpha) = \frac{\lambda}{2}+\frac{\omega^2}{2}\;\frac{\lambda}{(\alpha-\lambda)^2}.
\end{eqnarray}

\begin{figure}[t] 
\epsfxsize=\linewidth
\epsfbox{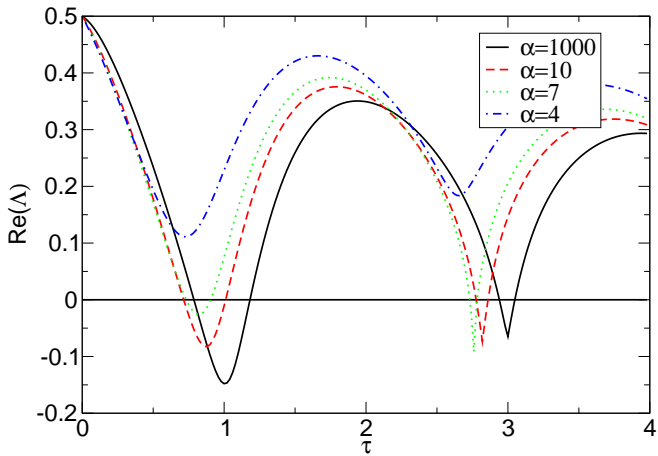}
\caption{\label{fig:figure8}(Color online) Largest real part of the complex eigenvalues $\Lambda$
vs. $\tau$ for $\lambda = 0.5$, $\omega = \pi$, and
$K=0.3$ as given by Eq.~(\ref{eqn:characteristic_low}). The solid, dashed, dotted, and dash-dotted curves correspond
to a cut-off frequency of $\alpha=1000$, $10$, $7$ and $4$, respectively.}
\end{figure}
The dependence of the minimal feedback gain $K_{min}$ on the cut-off frequency $\alpha$ is depicted in
Fig.~\ref{fig:figure7} for $\lambda=0.5$ and $\omega=\pi$. For large cut-off frequencies the minimum value tends to the
result of Sec.~\ref{sec:shape} [see Eq.~(\ref{eqn:K_min2})]. Note that for finite $\alpha$ the minimal feedback gain is
shifted to larger values compared to the case of the original TDAS control scheme.

\begin{figure}[t] 
\epsfxsize=1.0\linewidth
\epsfbox{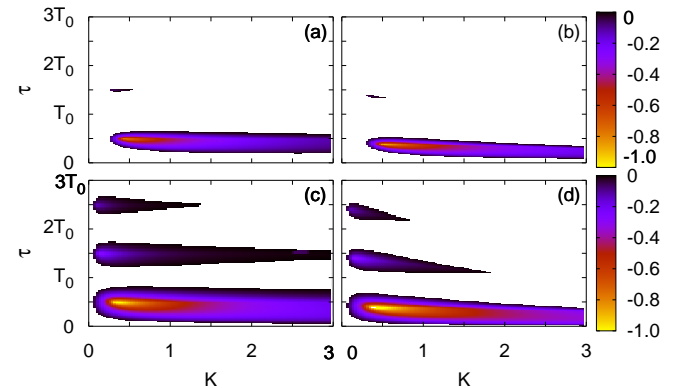}
\caption{\label{fig:figure9} (Color online)
Domain of control in the $K$-$\tau$~plane for different cut-off frequencies [panels (a) and (c): $\alpha=1000$, panels
(b) and (d): $\alpha=7$]: The shaded areas indicate combinations of $\tau$ and $K$, for which the largest real part
of the complex eigenvalues $\Lambda$ is negative and thus control is successful. The value of $\Re(\Lambda)$ is
indicated by the greyscale (color online). The parameters of the unstable focus are chosen as $\omega=\pi$ in all
panels and $\lambda=0.5$ in (a), (b) and $\lambda=0.1$ in (c), (d).}
\end{figure}
Figure~\ref{fig:figure8} shows the largest real part of the eigenvalues $\Lambda$ in dependence
on the time delay $\tau$ for fixed feedback gain $K=0.3$  and various cut-off frequencies $\alpha=1000$, $10$, $7$ and
$4$. For a large cut-off frequency, i.e., $\alpha=1000$, the curve is similar to the case without low-pass filter [see
Fig.~\ref{fig:figure2}(a)] indicating that the filter has only little effect. For smaller $\alpha$, however, filtering
the control signal reduces the range of the time delay $\tau$, for which $\Re(\Lambda)$ becomes negative, eventually
leading to a complete failure of stabilization. Note that the notches are shifted to lower values of $\tau$ for
decreasing $\alpha$. This effect can also be understood by Eq.~(\ref{eqn:tau_alpha}) due to the additional
factor $(1-\lambda/\alpha)<1$. The corresponding domains of control in the $K$-$\tau$~plane are shifted to smaller
values of $\tau$ and distorted asymmetrically (Fig.~\ref{fig:figure9}).

\section{Conclusion}
We have discussed the effects of time-delayed feedback control upon the stability of steady states. We have computed the
domain of stabilization of an unstable focus in the plane parametrized by feedback gain and time delay. Using the
complex multivalued Lambert function, we have derived analytically the main features of the stability domain by
investigating the characteristic equation of the fixed point. Below a minimum value of the feedback gain no control is
possible. In the vicinity of this minimum value, the shape of the domain of control shows a square root dependence on
the feedback gain. We find that no control is possible for time delays that are multiples of the uncontrolled
eigenperiod of the system. Taking non-zero control loop latencies into account, we have shown that increasing latency
times increase the minimum value of the feedback gain $K_{min}$ and reduce the domain of control substantially.
Similarily, an additional low-pass filter in the control loop causes a shift of $K_{min}$, as well. This suggests that
filtering with a cut-off frequency $\alpha$ has a similar effect as a latency delay time $\alpha^{-1}$. In fact,
expanding (\ref{eqn:K_min_latency}) for small latency $\delta$ in lowest order yields the same minimal feedback gain
$K_{min}$ as for low-pass filtering (\ref{eqn:K_min_lowpass}), if Eq. (\ref{eqn:tau_alpha}) is observed.

\section{Acknowledgement}
This work was supported by Deutsche Forschungsgemeinschaft in the framework
of Sfb 555. We are indebted to Andreas Amann and Wolfram Just for
stimulating discussions.


\begin{thebibliography}{10}
\bibitem{note1}
It should be noted that a similar characteristic equation as
Eq.~(\ref{eqn:characteristic}) holds for the Floquet exponents of a UPO,
where the lower bound, $K_{min} = \lambda/2$, of the feedback
gain has been shown to correspond to the flip threshold of control
\cite{JUS97,JUS99}.
\end{thebibliography}

\end{document}